\documentclass[a4paper,11pt]{article}
\pdfoutput=1 

\usepackage{jcappub} 

\usepackage[T1]{fontenc} 

\title{\boldmath The sinusoidal valley: a recipe for high peaks in the scalar and induced tensor spectra}

\author{A. Katsis}
\affiliation{National and Kapodistrian University of Athens,\\University  Campus,  Zographou,  157  84,  Greece}

\emailAdd{ariskatsis@phys.uoa.gr}

\abstract{Adding a sine-type interaction to inflationary models with two fields can evoke a classical trajectory with many turns in field space. Under conditions we discuss, the enhancement of the spectrum of adiabatic fluctuations resulting from each turn adds up. A special range of scales away from the CMB-constrained region can then be enhanced by several orders of magnitude, allowing for interesting phenomenological possibilities, such as induced gravitational waves or primordial black holes. A localized version of this interaction can also be used as an add-on to conventional inflationary models, thus allowing the injection of the large peak in their power spectra. The intuition and the conclusions drawn from this simple model remain relevant for more complicated applications that usually include extra terms that obscure the simplicity of the mechanism.}

\begin{document}
\maketitle
\flushbottom

\section{Introduction}

We interpret cosmological perturbations as resulting from the evolution of primordial, quantum fluctuations of the inflaton field(s)\cite{Guth:1982ec,Mukhanov:1981xt,Mukhanov:1990me}. Then working within the framework of cosmic inflation we can extract  information about the early universe by tracing back the evolution of late-time snapshots, like the observed CMB sky and the galaxy surveys for large-scale structure. For the simplest single-field slow-roll inflation models we obtain a scale invariant primordial scalar power spectrum that is fixed by the CMB-measured amplitude, $A_s \approx 2.2  \cdot 10^{-9}$ \cite{Planck:2018nkj,Planck:2018jri}. For the purposes of phenomenology and information extraction, a scalar spectrum of fluctuations that is enhanced for scales much smaller than the pivot scale of CMB, where it is not constrained by the large(r)-scale observations, is much more interesting.

Changes in the power spectrum of fluctuations, even dramatic ones like magnification by orders of magnitude \cite{Slosar:2019gvt}, can be attributed to exotic features of the background over which the fluctuations evolve. Various inflaton potentials inducing such an effect have been discussed in the literature \cite{Garcia-Bellido:1996mdl,Ivanov:1994pa,Garcia-Bellido:2017mdw,Silk:1986vc,Garcia-Bellido:2017fdg,Inomata:2021tpx,Dalianis:2021iig,Mukhanov:1997fw,Polarski:1994rz,Gordon:2000hv,Langlois:1999dw,Palma:2020ejf,Fumagalli:2020nvq,Pi:2017gih,Chen:2011zf,Fumagalli:2020adf}. In this paper we will construct inflationary models that result in a bump at the small scales of the scalar power spectrum and hence an expected increase in the power spectrum of the gravitational waves. Improving our ability to engineer such models will be beneficial in the quest of deciphering the signal of present and near-future detectors \cite{Hobbs:2009yy,LISA:2017pwj}. Furthermore, the enhanced scalar fluctuations that induce the gravitational waves have interesting implications on their own right, as they can collapse into primordial black holes \cite{Carr:1974nx,Carr:2020xqk,Chapline:1975ojl}. This speculative, yet well motivated form of matter could constitute (a part of) dark matter.

One of the recent theoretical advances in the field is the proposed mechanism that enhances the power spectrum (of adiabatic modes) in models of multifield inflation by exploiting its interaction with the power spectrum of isocurvature modes. The enhancement is induced by the so called `sharp turn in field space' 
\cite{Konieczka:2014zja,Achucarro:2010jv,Christodoulidis:2023eiw,Cespedes:2012hu,Achucarro:2012fd,Shiu:2011qw,Fumagalli:2021mpc,Spanos:2021hpk,Kallosh:2013yoa,Braglia:2020eai,Aldabergenov:2020bpt,LISACosmologyWorkingGroup:2024hsc,Fumagalli:2023loc,Kawai:2022emp}.
 In order to obtain a significant enhancement in the power spectrum, a very sharp turn has to take place. To arrange for that, one usually has to consider inflaton fields living  in an internal space with curved geometry or even consider non-intuitive potentials with polynomials in the denominator.
 More recently  it was argued that the same outcome is possible with a series of less sharp turns, without requiring inner curvature, by evoking what looks like a resonance effect \cite{Kefala:2020xsx}. The analysis of \cite{Boutivas:2022qtl} was based on analytic expressions for (the second slow-roll parameter) $\eta$ with several large pulses introduced by hand. We will show how one can devise a simple model that produces similar expressions as background inflationary solutions. 

In section \ref{sec:part1} we review turns in field space and assisted enhancement. In section \ref{sec:sinu} we construct the `sinusoidal valley' inflationary  potential and we exhibit how the mechanism results into a peak at small length scales of the scalar power spectrum. In section \ref{sec:part3} we show how a sinusoidal valley can be injected into other models, such as generalizations of conventional single-field inflationary models. A model-building recipe is provided in section \ref{sec:adv}, where the location and size of the peak in the spectrum are reverse engineered. At the same time, we argue how the sinusoidal valley can be connected with primordial back hole and gravitational wave observations. Finally, we discuss the current status of the physical motivation for sinusoidal valleys in section \ref{sec:disc}. In order to conform the numerical analysis, we find it convenient to express all dimensionful quantities in units of $M_{PL}$, but we suppress the units in the formulas for notational simplicity. Moreover, we have fixed the scale factor to be equal to $1$ today. In our convention $N=0$ corresponds to present time, inflation ends at (roughly) $N \approx -60$ and CMB exits the horizon at $N \approx -120$. For the power spectrum, the correspondence between the units is: $1 Mpc^{-1} = 2.63 \cdot 10^{-57} M_{PL}$.

\section{Assisted enhancement by turns in field space}\label{sec:part1}

\subsection{Classical evolution for turns in field space}\label{sec:turns}
Let us summarize the important elements of the evolution of the two-field system.  More details are provided in Appendix \ref{App:bg}.
We consider a simple action\footnote{Also used by \cite{Cespedes:2012hu}, but now we view it in the limit of trivial geometry in field space, namely when $\gamma_{ab} =\delta_{ab}$.} of two scalar fields, $\vec{\phi} =(\chi,\psi)$, with canonical kinetic terms, interacting through the potential $V(\chi, \psi)$:

\begin{equation}\label{equ:action}
S= \int d^4 x \sqrt{-g}\left[\frac{R}{2}-\frac{1}{2}g^{\mu \nu} \partial_{\mu} \chi \partial_{\nu}\chi-\frac{1}{2}g^{\mu \nu} \partial_{\mu} \psi \partial_{\nu}\psi  -V(\chi,\psi)\right]  \ .
\end{equation}
The gravitational background is a flat FRW metric, with line element:

\begin{equation}
ds^2 =-dt^2+a^2(t)\delta_{ij}dx^i dx^j \ ,
\end{equation}
where $a(t)$ is the standard scale factor describing the expansion of spatial surfaces. The background evolution is determined in terms of the background fields $\chi(t),  \psi(t)$ and $a(t)$.  For most of our analysis it will be convenient to measure time in terms of efolds of expansion, $dN=H(t)dt$ . We also rotate the field-space axes and parametrize with respect to the directions tangent and normal to the trajectory in field space \cite{Gordon:2000hv}.
The equations of motion read
\begin{subequations}\label{equ:eom}
\begin{align}
\label{equ:eom:1}
\chi_{,NN}+\left(3+\frac{H_{,N}}{H}\right)\chi_{,N}+\frac{V_{,\chi}}{H^2} & = 0 
\\
\label{equ:eom:2}
\psi_{,NN}+\left(3+\frac{H_{,N}}{H}\right)\psi_{,N}+\frac{V_{,\chi}}{H^2} &= 0 
\\
\label{equ:eom:3}
\frac{2V}{6-(\chi_{,N}^2+\psi_{,N}^2)} &= H^2 \ .
\end{align}
\end{subequations}
We decompose $\eta$ into its projections along these newly defined orientations to find

\begin{equation}\label{equ:eps}
\epsilon = \frac{1}{2}(\chi^2_{,N}+\psi^2_{,N}) \ ,
\end{equation}
\begin{equation}
\eta_T = 3+\frac{V_{,\chi} \chi_{,N}+V_{,\psi}  \psi_{,N}}{H^2(\chi^2_{,N}+\psi^2_{,N})} \ ,
\end{equation}
\begin{equation}\label{equ:etan}
\eta_N = \frac{V_{,\chi} \psi_{,N}-V_{,\psi} \chi_{,N}}{H^2(\chi^2_{,N}+\psi^2_{,N})} \ .
\end{equation}
$\eta_T$ is equivalent to the standard second slow-roll parameter of single-field inflation, while $\eta_N$ parametrizes the rate of turn of the trajectory in field space.

Finally, we consider perturbations from the classical trajectory $\vec{\phi}+\vec{\delta\phi}=(\chi,\psi)+(\delta\chi,\delta\psi)$, which we parametrize in terms of the curvature and isocurvature fields, $\mathcal{R}$ and $\mathcal{S}$  as follows:
\begin{equation}
\mathcal{R}=\frac{H}{\sqrt{\dot\chi^2+\dot\psi^2}}\left( \vec{T}\cdot \vec{\delta\phi} + \Psi \right) \ , \quad \mathcal{F}= \left( \vec{N}\cdot \vec{\delta\phi} \right) \ ,
\end{equation}
where $\mathcal{F}= \sqrt{2 \epsilon} \mathcal{S}$ is the rescaled entropy field \cite{Gordon:2000hv,Cespedes:2012hu,Achucarro:2012sm}, and $\Psi$ is the scalar part of the spatial metric perturbation required to form the Mukhanov-Sasaki perturbations. In order to determine the spectrum of the fluctuations in two-field inflation we need to solve the following  system for the coupled evolution

\begin{subequations}\label{equ:flu}
\begin{align}
\label{equ:flu:1}
\mathcal{R}_{k,NN}+r_1(N) \mathcal{R}_{k,N}+r_2(N)k^2\mathcal{R}_{k} &= -r_3(N)F_{k,N}-r_4(N)F_k \ ,
\\
\label{equ:flu:2}
\mathcal{F}_{k,NN}+s_1(N) \mathcal{F}_{k,N}+\left(s_4(N)+r_2(N)k^2\right)\mathcal{F}_{k}&= s_3(N)R_{k,N}  \ ,
\end{align}
\end{subequations}
where $\mathcal{R}_k$ and $\mathcal{F}_k$ are the Fourier-transformed quantities and everything is written in terms of efolds $N$. If one assumes that the evolution starts in the Bunch-Davies vacuum, the standard initial conditions to impose are:

\begin{equation}\label{equ:icBD}
\mathcal{R}_{ini}= e^{-\frac{3 N}{2}}\left(J\left(\frac{3}{2}, e^{-N} \frac{k}{H}\right) + i \  J\left(-\frac{3}{2}, e^{-N} \frac{k}{H}\right)\right) \ , \quad \mathcal{F}_{ini}= \sqrt{2 \epsilon} \  \mathcal{R}_{ini} \ .
\end{equation}
The power distributed to each mode can be deduced by squaring the amplitude of the corresponding fluctuation, 

\begin{subequations}\label{equ:PS}
\begin{align}
P_{\mathcal{R}} &= \frac{k^3}{2 \pi^2}  |\mathcal{R}_k |^2 \ ,
\\
P_{\mathcal{F}} &= \frac{k^3}{2 \pi^2}  |\mathcal{F}_k |^2  \ .
\end{align}
\end{subequations}
The explicit formulas for the evolution coefficients are:

\begin{subequations}
\begin{align}
r_1 \equiv 3+\epsilon-2 \eta_T \ , \quad r_2 \equiv \frac{e^{-2N}}{H^2} \ , \quad r_3 &\equiv 2 \frac{\eta_N}{ \sqrt{2 \epsilon}} \ , \quad r_4 \equiv 2 \frac{\eta_N}{\sqrt{2 \epsilon}}\left(3-\eta_T+\frac{\eta_{N,N}}{\eta_N}\right)
\\
s_1  \equiv 3-\epsilon \ , \quad s_3 &\equiv 2  \sqrt{2 \epsilon} \eta_N \ , \quad s_4 \equiv   (\mu/H)^2 - \eta_N^2  \  ,
\end{align}
\end{subequations}
and we have also defined the bare mass of the fluctuation
\begin{equation}\label{equ:M}
\mu^2 \equiv N_{\chi}N_{\chi}V_{, \chi \chi} +N_{\psi}N_{\psi}V_{, \psi \psi}+2N_{\chi}N_{\psi}V_{, \chi \psi} \ .
\end{equation}
If one expands the action to quadratic order in these fluctuations, the coefficient of the $\mathcal{F}^2$ term, which we usually define as effective mass of the fluctuation, turns out to be \cite{Cespedes:2012hu} :
\begin{equation}
M_{eff}^2 \equiv \mu^2 - \frac{N_{\chi}V_{, \chi }+N_{\psi}V_{, \psi }}{\sqrt{\dot\chi^2+\dot\psi^2}}=\mu^2-(\eta_N H)^2 = s_4 H^2 \ .
\end{equation}

\subsection{A mechanism that enhances fluctuations}\label{sec:fluct}

The system \eqref{equ:flu} is precisely what we mean when we say that features of the background can get imprinted into the spectra of fluctuations. The coefficients that control the evolution are the background quantities. For a single field we have $\mathcal{F}=0$ and only the first equation survives

\begin{equation}\label{equ:step}
\mathcal{R}_{k,NN}+(3+\epsilon-2 \eta_T)\mathcal{R}_{k,N}+\frac{e^{-2N}}{H^2}   k^2\mathcal{R}_{k}=0 \ .
\end{equation}
When the background evolution involves steps in the inflaton potential or inflection points, the resulting pulses of $\eta_T$ are translated into a temporary negative coefficient for $\mathcal{R}_{k,N}$ in eq. \eqref{equ:step} and thus a corresponding increase of the amplitude of $\mathcal{R}_k$.

The coupled evolution reveals another interesting possibility, the enhancement of curvature perturbations assisted  by a temporary increase in the isocurvature perturbations. Let us briefly sketch how we expect this to work. The centrifugal barrier that originates from the rotation of the trajectory reduces the effective mass of the isocurvature mode, as we see in eq. \eqref{equ:M}. When $(\mu/H)^2 \gg \eta_N^2$, the coefficient $s_4$ will be positive and dominant in the evolution of the isocurvature modes, resulting into a quick suppression of $\mathcal{F}_k$. On the contrary, for $(\mu/H)^2 \ll \eta_N^2$  the centrifugal barrier renders the effective mass of the isocurvature fluctuation temporarily negative, or equivalently  $s_4<0$. As a result, $\mathcal{F}_k$ increases. At the same time, sufficiently large $\eta_N$ affects the values of  $r_3$ and $r_4$. When turned on, $\mathcal{F}_k$ acts as a source for the increase of $\mathcal{R}_k$.\footnote{Let us note that this argument involves the \`a la carte treatment of eq. \eqref{equ:flu} as a system and as separate equations and is thus only  indicative of the expected evolution. As we will show in the following, it still captures the core of the mechanism.} If further turns occur before $\mathcal{F}_k$ has time to return to its previous value, then the isocurvature and the adiabatic amplitude can gradually increase by several orders of magnitude. When the turns are over,  $\mathcal{F}_k$ loses its support from $s_4$ and sharply decays. Then the evolution of the adiabatic amplitude is again described by eq. \eqref{equ:step}, reproducing the well-known constant superhorizon solution. The difference here is that $\mathcal{R}_k$ freezes at the enhanced value.

We conclude that in order to get a significant amount of `assisted enhancement' we require a period of time where frequent, sharp turns in field space satisfying 
\begin{equation}\label{equ:cond0}
|\eta_N| > 0 \ 
\end{equation} 
take place, such that the effective isocurvature mass is destabilized: \footnote{Our $Q$ is equivalent to  $(1- \xi) \eta_N^2$ of ref. \cite{Fumagalli:2020nvq}.}
\begin{equation}\label{equ:cond}
Q \equiv   (\eta_N)^2- (\mu/H)^2=-(M_{eff}/H)^2 > 0 \ .
\end{equation}

\section{The sinusoidal valley potential}\label{sec:sinu}

\subsection{Crafting a turning trajectory}

\begin{figure}[t]
\centering
\includegraphics[width=0.5\textwidth]{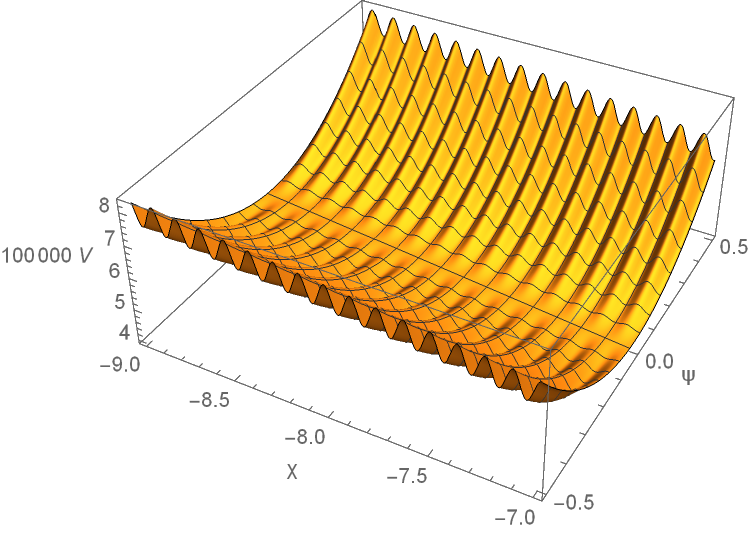}
\includegraphics[width=0.4\textwidth]{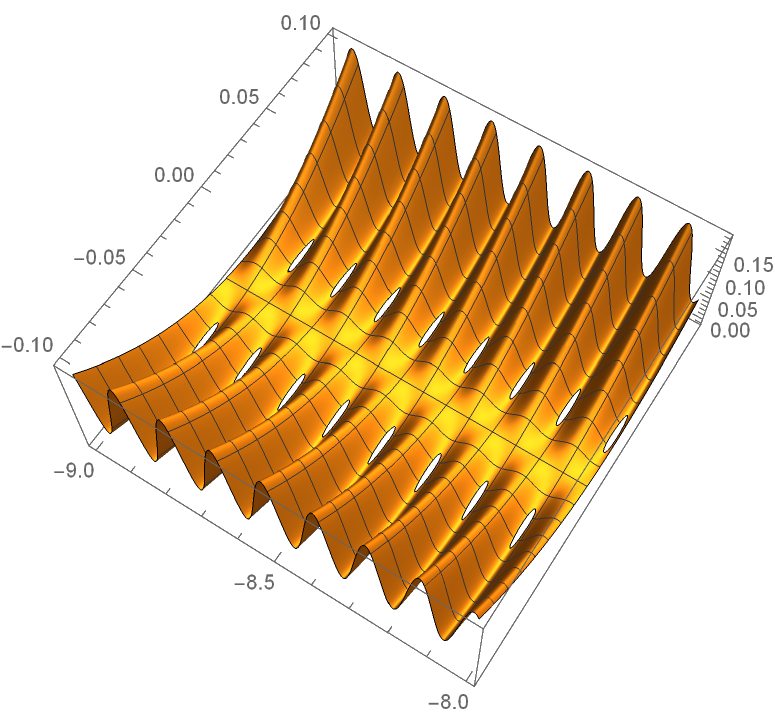}
\caption{
Left: a segment of the potential $V(\chi,\psi)$ of eq. \eqref{equ:pot1} for the choice \eqref{equ:model0}. Right: the local minima (wells) dug by the sine term into the valley in grey.
}
\label{fig:pot1}
\end{figure}

The background evolution, eq. \eqref{equ:eom} or equivalently eq. \eqref{equ:eomt}, bears resemblance to the classical problem of a ball traveling across a mountainous territory of height $V(\chi,\psi)$ at each point ($\chi$,$\psi$), rolling down the hills under the influence of gravity and moving freely along the valleys. Additionally, the terrain has friction, represented by the Hubble parameter. This allows us to engineer a toy model, the background evolution of which features turns in field space. The idea is to use a potential that approximately constrains the classical motion to a turning trajectory in ($\chi ,\psi$). The sine is an obvious choice for a function with several turns. If the potential energy scales  like $\sim \psi\left(\psi -  \sin( \chi)\right)$, then the trajectory $\psi \sim  \sin( \chi)$ will be favored by energy arguments. Our choice  \footnote{In similar fashion, one could constrain the motion to the turning trajectory more tightly by using a square $\left(\psi -  \sin( \chi)\right)^2$, an exponential or even a delta function in the potential.} is the toy-model potential
\begin{equation}\label{equ:pot1}
V(\chi,\psi)=  \psi \big( m^2 \psi - b  \sin( \omega \chi) \big) - c \ \chi  \ .
\end{equation}
The primary feature of this potential is the sinusoidal valley around $\psi=0$, which we depict in figure \ref{fig:pot1}. The walls that encode the extra energy cost to deviate from the valley trajectory are approximately quadratic and their steepness depends on $m^2$. The valley has an overall constant tilt $c$ on the $\chi$-axis, so that when we place a ball at a given point of it, we expect it will snake its way down, along the valley.   The sine term digs shallow wells into the valley of the potential when $ \psi\big(m^2 \psi - b  \sin(\omega \chi)\big)$ takes its minimum (negative) value. The wells deflect the motion of the inflaton, but can also trap the inflaton if the tilt  $c$ is too small.

We will actually apply the background analysis to a modified version of the model which is more suitable for phenomenology
\begin{equation}\label{equ:pot1bm}
V(\chi,\psi)=  \psi \big( m^2 \psi - b \ W  \sin( \omega \chi) \big) - c \ \chi +  V_m  \ ,
\end{equation}
where we have introduced a modification of the last stage of inflation
\begin{equation}\label{equ:Vm}
V_m=  c \big(1 + \tanh(\chi) \big)\chi   \ ,
\end{equation}
so that a minimum is formed, where reheating will take place after the end of inflation. This late modification does not affect the earlier stages of inflationary evolution.

\begin{figure}
\centering
\includegraphics[width=0.5\textwidth]{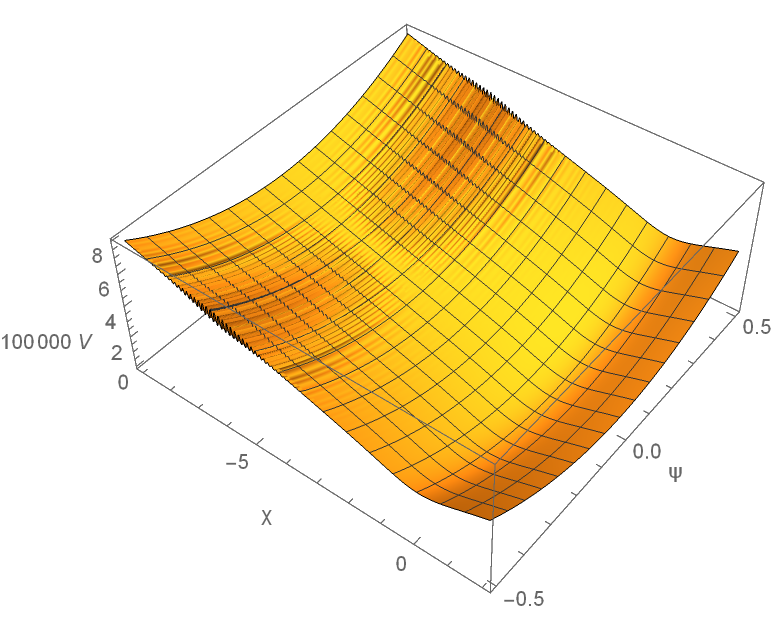}
\caption{The modified potential $V(\chi,\psi)+V_m(\chi,\psi)$ of eqs.  \eqref{equ:pot1bm}-\eqref{equ:We} that describes a sinusoidal valley inside the envelop of a potential with a global minimum.}
\label{fig:pot2m}
\end{figure}
We have also introduced an envelope on the sine-type interaction that causes the turns in field space,
\begin{equation}\label{equ:We}
W(\chi)=1-\tanh^2\big(\rho(\chi-\chi_*)\big) \ .
\end{equation}
This keeps the mechanism of assisted enhancement exclusive to the central part of the potential around $\chi=\chi_*$.
Generically, we expect from eq. \eqref{equ:cond} that the  turns in field space in sinusoidal valleys become gradually more frequent and pronounced as  $H$ decreases  at later times. Most of the enhancement would then occur right before reheating, which is outside the scope of this paper.\footnote{A naive extrapolation of our calculation would anyway predict huge pulses of $Q$ at the minimum but since the modified dynamics of reheating have not been taken into account this analysis is not valid.} If we take care that the final part of the inflationary evolution does not feature significant turns in field space, then the modes can freeze at their enhanced amplitude before the inflaton reaches the minimum. As a bonus, from a numerical point of view, such a choice allows us to motivate (Bunch-Davies) initial conditions for the modes of interest.  The potential of model \eqref{equ:pot1}-\eqref{equ:model0} is depicted in figure \ref{fig:pot2m}.

Let us now confirm our expectations for the trajectory in field space, by computing numerically.  We pick the following parameter values
\begin{equation}\label{equ:model0}
\omega =50 \ ,  \  c= 5.7 \cdot 10^{-6} \ , \ m^2=1.14 \cdot 10^{-4} \ , \ b=7.17 \cdot 10^{-6} \ , \  \chi_*=-6 \ ,  \ \rho=0.9 \ ,
\end{equation}
and we integrate the equations of motion from $N_i=-127$ to $N_f=-70$. It would be meaningless to compute beyond $N_f$, because the system of equations \eqref{equ:flu} does not capture the physics of reheating or the subsequent evolution stages. We start the evolution a few efolds earlier at a sensible part of the potential, say at $(\tilde{\chi}_{ini},\tilde{\psi}_{ini})=(-9,0)$, and then record as initial conditions the numbers we get for $N=-127$. We find $ (\chi_{ini},\psi_{ini})=(-8.74327,0.00121)$ and $(\dot{\chi}_{ini},\dot{\psi}_{ini})=(0.11375,-0.00144)$. 
\begin{figure}[t]
\centering
\includegraphics[width=0.45\textwidth]{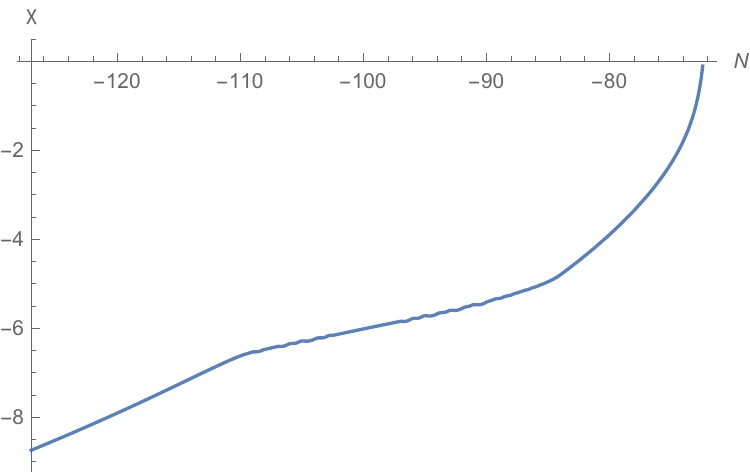}
\includegraphics[width=0.45\textwidth]{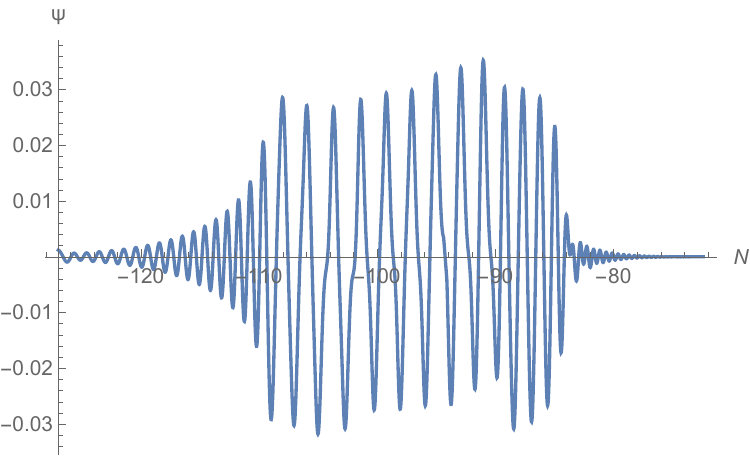}
\caption{
The resulting oscillating background solutions for the scalar fields $\chi(N)$ and  $\psi(N)$  for the model \eqref{equ:pot1} with the choice of parameter values \eqref{equ:model0}.
}
\label{fig:xy1}
\end{figure}

\begin{figure}[t]
\centering
\includegraphics[width=0.45\textwidth]{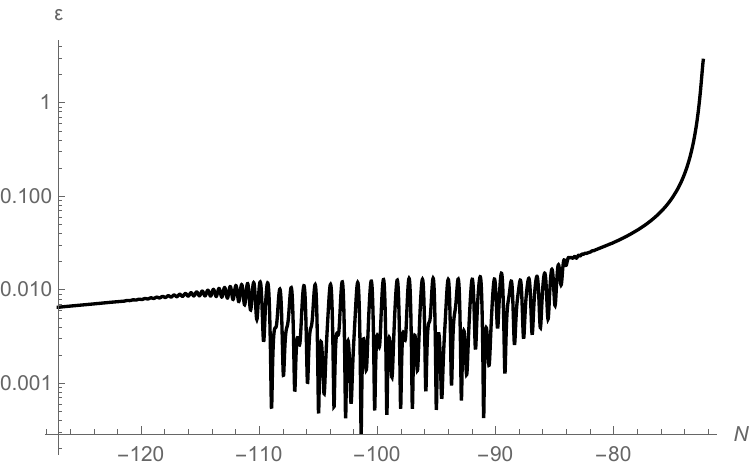}
\includegraphics[width=0.45\textwidth]{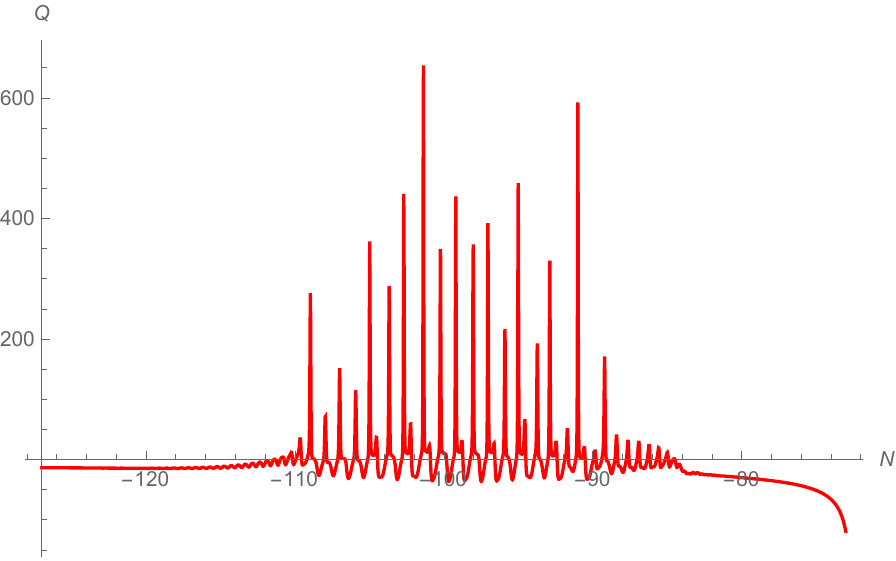}

\includegraphics[width=0.45\textwidth]{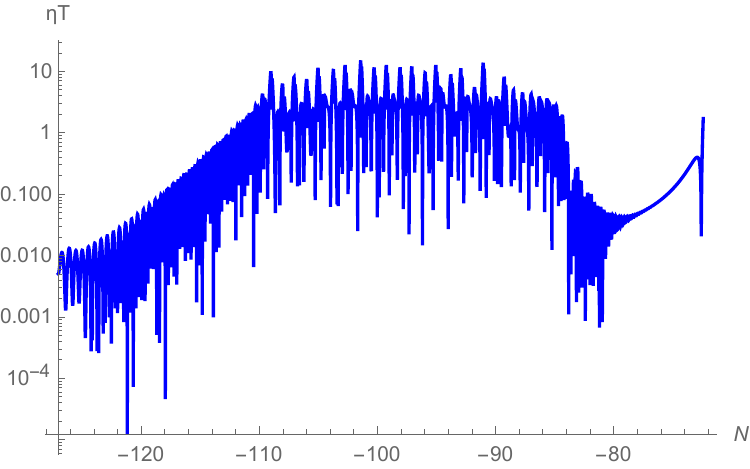}
\includegraphics[width=0.45\textwidth]{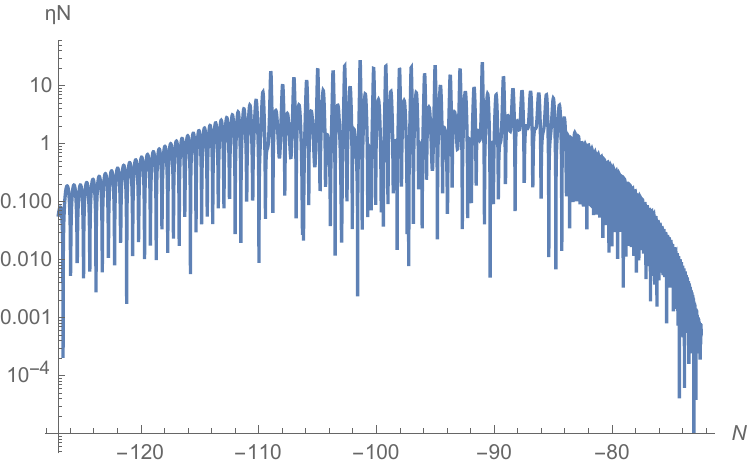}
\caption{
For the numerical solution of the background evolution for \eqref{equ:pot1}  and  \eqref{equ:model0}: the slow-roll parameters $\epsilon(N)$ (black), $|\eta_T(N)|$ and $|\eta_N(N)|$ (blue) and the quantity $Q$ of eq. \eqref{equ:cond} (red).
}
\label{fig:sr1}
\end{figure}
The numerical solution (figure \ref{fig:xy1}) indeed features oscillations in $\psi$ at the central part.  In addition to the expected snaking motion, there are also small oscillations in $\chi$. 
The results for $\epsilon$, $|\eta_T|$,  $|\eta_N|$ and $Q$  are depicted in figure \ref{fig:sr1}. Had we not introduced the envelope, the oscillations of all plotted quantities would cover the whole time-span.

 \subsection{Arranging for sharp and destabilizing turns}
 
\begin{figure}[t]
\centering
\includegraphics[width=0.6\textwidth]{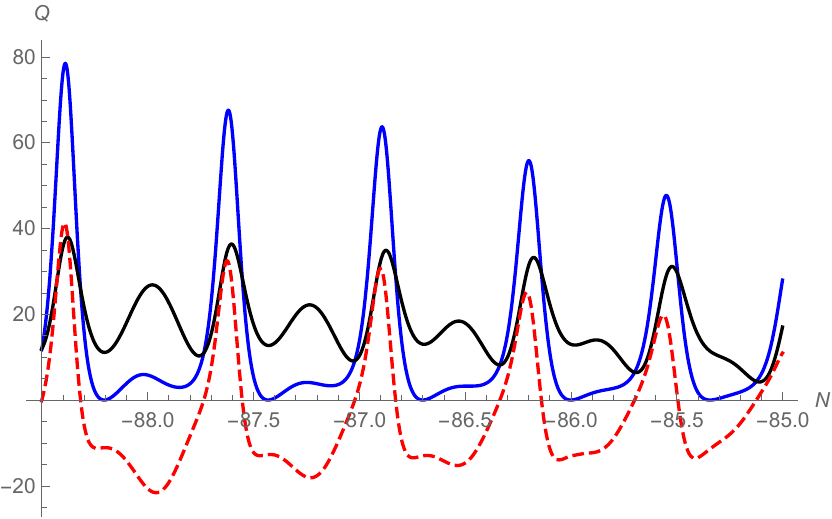}
\caption{
For the numerical solution of the background evolution for \eqref{equ:pot1} and \eqref{equ:model0}, we zoom into a shorter timespan and plot: a) in black the quantity $(\mu/H)$, b) in blue the quantity $|\eta_N|$ and c) in red dashed the resulting pulses of positive $Q$, as defined in eq. \eqref{equ:cond}. 
}
\label{fig:fun}
\end{figure}

The sinusoidal valley potential explicitly engineers the turns in field space, but the condition eq. \eqref{equ:cond} is not necessarily satisfied. In fact, a different choice of parameters can yield $Q<0$ for the whole trajectory. As we show in figure  \ref{fig:fun}, the quantity $(\mu/H)^2$ is  bigger than $\eta_N^2$ for most of the trajectory and it decreases only in places where no peaks in $\eta_N$ occur. The peaks of $\eta_N^2$ due to the turns have to be higher than $(\mu/H)^2$ to allow for pulses of positive $Q$. 

According to the analytical considerations that we present in Appendix \ref{App:anal}, pulses of positive $Q$ are obtained when the $\chi$-axis motion slows down to a near-stop at the max $\psi$-axis-displacement points, which are the turning points of the sinusoidal valley.
Keeping the rest of the parameters fixed, increasing $b$ results into more efficient slowdown and gradually sharper turns. For $b$ greater than some value $b_{crit}$, the inflaton gets trapped in a local minimum. We can approximate that:

\begin{equation}\label{equ:csd}
b_{crit} = 2m \sqrt{\frac{( c -\delta)  }{ \omega}}   \approx 2 m\sqrt{\frac{c  }{ \omega}} \ ,
\end{equation}
where $\delta$ is a small number.  We will refer to this as the `critical slow-down' approximation. For our discussed model of \eqref{equ:pot1} and \eqref{equ:model0}, it is $b_{crit} \approx 7.1 \cdot 10^{-6}$. In contrast, for $b=4.5 \cdot 10^{-6}$ we find that $Q<0$ for the whole trajectory.

Inspecting figure \ref{fig:ex4Pk}, one notices that the height of the coloured lines is in correspondence when we plot the power spectrum  in logarithmic axis and $Q$ in a normal axis. The peak enhancement thus appears to scale as $P \sim P_0\exp{(Q)}$. The spectrum is indeed modelled as a such exponential of the turning parameter in \cite{Boutivas:2022qtl} and \cite{Fumagalli:2020adf}. The dependence of $Q$ on $b$ is  harder to deduce. As we roughly approximate in \eqref{equ:expb}, it could also be depending on the relative difference of $b$ from its critical value exponentially. The resulting peak enhancement would then  scale as a double exponential of this difference, $P \sim P_0 \exp{[ Q_0 \exp{\left(-c_1|b_{crit}-b|/b_{crit})\right)}}]$. All findings in this paper suggest that the enhancement decreases very fast as the relative difference of $b$ from the critical-slow-down value increases, and a $40\%$ difference is usually more than enough to stop any enhancement.

\subsection{Assisted enhancement of the adiabatic amplitude}\label{sec:enh}

 \begin{figure}[t]
\includegraphics[width=0.45\textwidth]{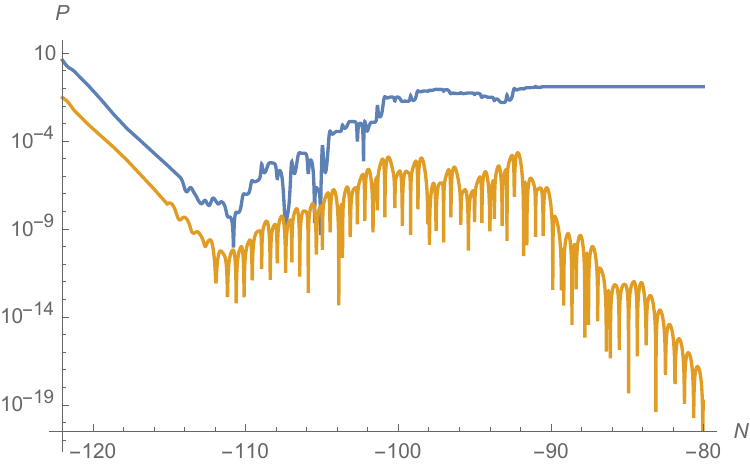}
\includegraphics[width=0.45\textwidth]{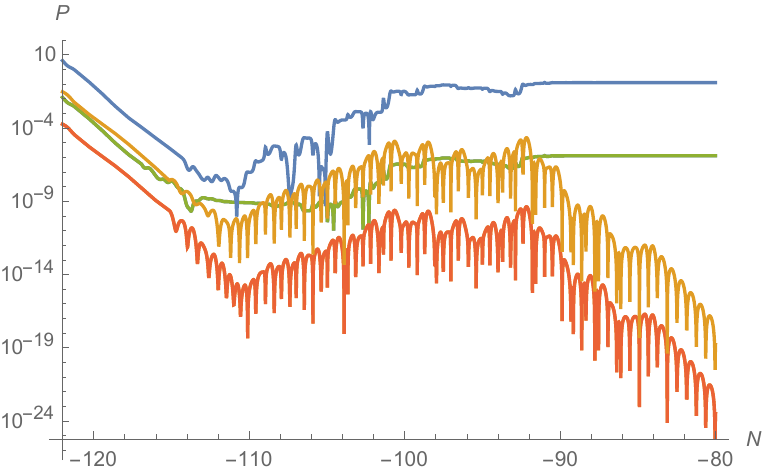}
\caption{For the model \eqref{equ:pot1bm} to \eqref{equ:model0}, Left: for the Fourier mode $k_1=1.8 \cdot 10^{-51}$ we plot $P_R(N)$ (blue) and $P_F(N)$ (orange). Right:  the same plot, also depicting $k_2=10^{-52}$ in green and red.
}
\label{fig:ex3kP}
\end{figure}

Let us begin with our analytical expectations for the evolution of a single mode $k_1$ according to the system \eqref{equ:flu}. The initial stage of evolution corresponds to the period of time before the feature is crossed, $N \ll N_*$, where $N_*$ is the time when the center of the feature is crossed so that $\chi(N_*)=\chi_*$. Since $\eta_N=0$, the adiabatic amplitude is decoupled from the isocurvature amplitude and they evolve approximately as

\begin{equation}\label{equ:kvk}
\mathcal{F}_{k,NN}+3 \mathcal{F}_{k,N}+\left((\mu/H)^2  + \frac{k^2 e^{-2N}}{H^2}\right)\mathcal{F}_{k}= 0  \ ,
\end{equation}
\begin{equation}\label{equ:kvk2}
\mathcal{R}_{k,NN}+3 \mathcal{R}_{k,N}+ \frac{k^2 e^{-2N}}{H^2} \mathcal{R}_{k}= 0  \ .
\end{equation}
This yields an increasingly\footnote{The decay is amplified with time because $(a H)^{-1}$ is a decreasing function of time during inflation.} decaying solution for both amplitudes.  During the final stage of evolution, $N \gg N_*$, the last term receives an exponential suppression, approximately leaving us with

\begin{equation}\label{equ:kvk3}
\mathcal{F}_{k,NN}+3 \mathcal{F}_{k,N}+(\mu/H)^2 \mathcal{F}_{k}= 0  \ ,
\end{equation}
\begin{equation}\label{equ:kvk4}
\mathcal{R}_{k,NN}+3 \mathcal{R}_{k,N} = 0 \ .
\end{equation}
Consequently, the isocurvature amplitude decays while the adiabatic amplitude acquires a constant mode.
The intermediate stage of evolution around $N_*$ is where we expect the assisted enhancement to take place. Notice that $r_2$ is the only term that tells different $k$ apart in eqs. \eqref{equ:flu}. During the second half of the enhancing stage, this term becomes subleading and all wavenumbers experience the same evolution.

Next we solve the coupled system \eqref{equ:flu} for the fluctuations $\mathcal{R}_k$ and $\mathcal{F}_k$ using the initial conditions \eqref{equ:icBD} to confirm numerically the mechanism of assisted enhancement of fluctuations, as sketched in section \ref{sec:fluct}.
We show the evolution of the scalar amplitude for the mode  $k_1= 1.8 \cdot 10^{-51}$ for $-122 \leq N \leq -80$ in the first plot of figure \ref{fig:ex3kP}. The adiabatic mode is in blue and  the isocurvature mode in orange. During the intermediate stage, $-110 \leq N \leq -90$, there are large pulses of positive $Q$ corresponding to the sharp turns in field space. We confirm that these pulses pump the isocurature amplitude up one after the other. For each oscillation in $|\eta_N|$, i.e. each turn, there is a corresponding oscillation in the isocurvature amplitude (yellow). Initially, large turns occur very frequently, the isocurvature amplitude has no time to die out between the pulses and ends up increasing by many orders of magnitude in several small steps. This is the resonance phenomenon discussed in \cite{Kefala:2020xsx}.  During each turn,  the coupling with the adiabatic amplitude is turned on and the isocurvature sources the enhancement of the adiabatic amplitude in a step-by-step procedure. The total enhancement is roughly $7.5$ orders of magnitude  compared to the CMB value. The temporary absence of turns after $N=-90$ leads to a quick suppression of the isocurvature amplitude and the assisted enhancement ceases. The adiabatic amplitude soon freezes at its enhanced value.

\subsection{The enhanced power spectrum}\label{sec:enh2}

\begin{figure}[t]
\centering
\includegraphics[width=0.45\textwidth]{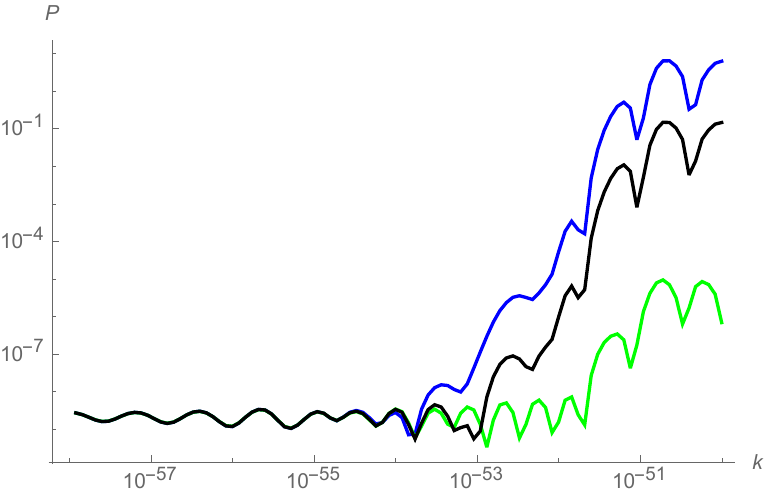}
\includegraphics[width=0.45\textwidth]{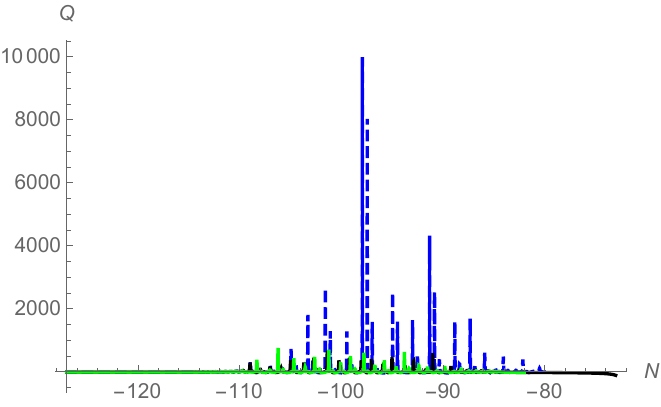}
\caption{For the model \eqref{equ:pot1bm} to \eqref{equ:model0} we plot, left: in black ($b=7.17 \cdot 10^{-6}$), the scalar power spectrum of the adiabatic mode, $P_R(k)$, for the band of Fourier modes $10^{-58} \leq k \leq  10^{-50}$ at $N=-70$, when the modes have frozen; in green and in blue, the same quantity for $b=7 \cdot 10^{-6}$ (green) and $b=7.19 \cdot 10^{-6}$ (blue). Right : the corresponding pulses of positive $Q$ for the three choices of $b$ with the corresponding colors.  
}
\label{fig:ex0k3}
\end{figure}

The power spectrum depends on the magnitude of different $\mathcal{R}_k$ modes. Let us  we consider the evolution of a second mode $k_2<k_1$. 
We begin with the initial stage of evolution, $N<N_*$. From eq. \eqref{equ:kvk} we can deduce that $\mathcal{F}_{k2}$ drops faster. If we ignore the relatively small change in $H$, then the explicit time dependence of the evolution equation is encoded in $(k/a)^2$. We could equivalently view $k_2$ as following the same evolution equation as $k_1$ but being slightly forward in time, $\tilde{N}=N+\log(k_1/k_2)$. This expectation is confirmed for $k_2 = 10^{-52}$ in figure \ref{fig:ex0k3} when $N<-114$, where the different amplitudes follow predictably different evolution. The scalar amplitude for the adiabatic mode $\mathcal{R}_{k2}$ is depicted in the second plot in green. We also show the corresponding quantity for the isocurvature mode $\mathcal{F}_{k2}$ in red. The red line would become parallel to the yellow line if we shifted it a few efolds forward in the time axis. They do not become identical because the initial conditions for them differ. 

During the first half of the enhancing stage, $-114 < N< -102$, the isocurvature amplitudes $\mathcal{F}_{k}$ already follow a similar evolution while the adiabatic amplitudes $\mathcal{R}_{k}$ are quite different. 
During the second half of the enhancing stage, all modes experience the same, `parallel' evolution, as expected. Namely, for $N>-102$ the red line is just a shifted down version of the yellow line. A similar behavior is evident for the adiabatic modes: $\mathcal{R}_{k2}$ is a shifted down version of  $\mathcal{R}_{k1}$. 
We note that the corresponding amplitudes of the modes $k_1$ and $k_2$ start with roughly $2.5$ orders of magnitude separating them and exit the enhancement with roughly $5$ orders of magnitude separating them. This deviation occurs gradually during the initial stage and mostly during the first half of the enhancing stage.

Now consider a much larger wavenumber, $k_0>>k_1$. This mode is deeply inside the horizon when the feature is crossed and the $k^2$ term is dominant in eqs. \eqref{equ:flu}. Consequently, it is  unaffected by the assisted enhancement and its adiabatic amplitude remains small. If we gradually decrease the wavenumber we will arrive at the scenario where only the final part of the feature is scanned when the mode is about to cross the horizon. 
Such a mode experiences a smaller number of turns in field space and consequently a milder enhancement compared to $\mathcal{R}_{k1}$. This is similar to the peak amplitude for the case where fewer turns in field space, or equivalently enhancement steps are present. 

We conclude that the power spectrum will peak at scale $k_{peak}$ that corresponds to wavenumbers that are about to cross the horizon when the start of the relevant feature is scanned, about $N_*$. Smaller wavenumbers enter the enhancement stage with suppressed amplitude while larger wavenumbers will skip some or even all of the enhancing stage, being deep inside the horizon when it takes place.  For $H$ approximately constant during the part of the inflationary evolution that features the turns in field space, a wavenumber $k$ crosses the Hubble horizon at $N=\log(k/H)$. Then we find
 \begin{equation}\label{equ:peakenh}
k_{peak} \approx H \exp(N_*) \ . 
\end{equation} 

 At the left panel of figure \ref{fig:ex0k3} we show the power spectrum of the adiabatic mode for $ 10^{-58} \leq k \leq  10^{-50}$. As expected, very small wavenumbers remain unaffected. For $b=7.17 \cdot 10^{-6}$ (in black), assisted enhancement results into a wiggly bump with peak amplitude of $0.15$ at about $k_{peak}=2.1 \cdot 10^{-51}$ .  
The enhancement of the spectrum is most sensitive to the value of $b$. The precision to which the value satisfies the critical slow-down can change the result by several orders of magnitude. In figure \ref{fig:ex0k3} we also plot the pulses of positive $Q$ and the corresponding spectra for  $b=7 \cdot 10^{-6}$ (green) and $b=7.19 \cdot 10^{-6}$ (blue). The number and size of pulses  of positive $Q$ increases as we go close to the critical slow down and results into greater enhancement. The maximum enhancement is also influenced by the thickness of the envelope, $\sim 1/\rho$. When keeping the rest of the parameters fixed, the enhancement of a wider envelope is stronger because it allows more consecutive turns in field space.

\section{Introducing a sinusoidal valley in inflationary models}\label{sec:part3}

\subsection{The sinusoidal valley as a universal add-on}\label{sec:gen}

In section  \ref{sec:sinu} we analyzed the simplest sinusoidal valley, the valley of constant average $\chi$-direction tilt. There are other types of valleys, with varying  tilt, that exhibit the same interesting features.  We can imagine a general sinusoidal valley as consisting of many strips. Each strip has constant average tilt but different than its neighboring strips, and is like a very short sinusoidal valley of the type we previously discussed. Then the inflaton will roll too fast on strips with greater tilt and will be trapped on strips with smaller tilt. To avoid that, we have to focus the envelope \eqref{equ:We} on the part of the valley where the inflaton has slowed down but does not get trapped. This will be the relevant part of the potential where the sinusoidal valley will appear and the assisted enhancement will take place. 

In fact, any single-field inflationary model can be generalized to include a sinusoidal valley. All one has to do is add a second, free and presumably heavy field and then couple it with the conventional inflaton using the temporary\footnote{We can view the envelope imposed on the interaction as a localization in $\chi$ field direction or equivalently in time.} sine-type interaction.  When the inflaton scans the sinusoidal valley, the mechanism of assisted enhancement will operate. The benefit of doing so is that  most of the predictions of the original single-field model, satisfying observational constraints, etc, are roughly unaffected and a bump is introduced in the power spectrum at larger wavenumbers on top of them. 
The steps to follow are summarized in the following `land-shaping' recipe:

\begin{itemize}

\item Pick a working single-field inflation model, whose parameter choices make it consistent with observations, say $\mathcal{L}_{sf}(\chi)$.

\item Introduce a second, massive inflaton field by adding its free Lagrangian, $\mathcal{L}_{free}(\psi)$. This trivially upgrades the theory to a 2d field space with a straight line valley potential of varying depth. For the standard initial conditions where the field starts about the valley we get the same trajectory in field space as in the single-field case. The predictions for the spectrum do not change.

\item Introduce the sine-type interaction, $\mathcal{L}_{sinu}(\chi,\psi)$. This twists the potential into the sinusoidal valley. For the case with the envelope, a sinusoidal valley potential appears somewhere inside the straight-line valley. The mechanism of assisted enhancement will add a bump in the scalar power spectrum, focused around the scale $k_{peak}$ given by eq. \eqref{equ:peakenh}.

\item Tune the value of $b$ (and the size of the envelope) to obtain larger enhancement by moving closer to the `critical-slow down' value, $b_{crit}$. In practice, it is easy to find this value as the maximum value of $b$ for which the classical trajectory ends on the true minimum rather than on an intermediate, local minimum.

\end{itemize}

\subsection{Example: valley of parabolic tilt}\label{sec:square}

Let us apply the recipe to the simple and favorite pedagogical model,  $m^2 \chi^2$ inflation,
\begin{equation}\label{equ:modelb1}
V(\chi_, \psi) =  V_{\chi}+V_{\psi} +V_{sinu} \ ,
\end{equation}
where he have added to the standard free field $\chi$
\begin{equation}\label{equ:modelxsq}
V_{\chi} =  m_{\chi}^2 \chi^2 \ ,
\end{equation}
a second free field $\psi$ 
\begin{equation}\label{equ:modelb2}
V_{\psi} =  m_{\psi}^2 \psi^2 \ ,
\end{equation}
and then the localized sine-type interaction between them
\begin{equation}\label{equ:modelb3}
V_{sinu}= -b W \psi   \sin( \omega \chi)  \ .
\end{equation}

Roughly speaking, the values used for the single-field evolution in $V_{\chi}$, namely $m_{\chi}$ and $\chi_i$, determine the spectral index and the power spectrum about the CMB scales. We pick  $m_{\chi}^2=2.3 \cdot 10^{-7}$ and $\chi_i=-14.56$. Then we  extend to a two-field model by introducing $V_{\psi}$. As long as the initial point of the trajectory is inside the valley, here $\psi_i=0$, and the $m_{\psi}$ is large enough, the background evolution of $\chi$ and the adiabatic scalar power spectrum remain identical to the single-field case. Introducing $V_{sinu}$ twists the valley into the sinusoidal shape that dictates the turns into field space. The parameters $b$ and $\omega$ determine the enhancement of the spectrum. The envelope parameter $\chi_*$ determines which scales will get enhanced. The other envelope parameter, $\rho$, affects the amount of enhancement by allowing for more or less turns about $\chi_*$.

\begin{figure}[t]
\centering
\includegraphics[width=0.5\textwidth]{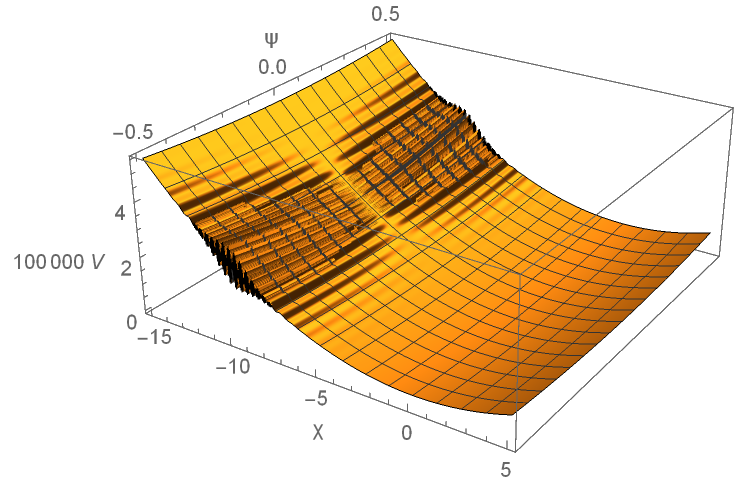}
\caption{
The sinusoidal potential for the model \eqref{equ:modelb1} to \eqref{equ:modelxsq1}.
}
\label{fig:potxsqtan}
\end{figure}

\begin{figure}[t]
\includegraphics[width=0.45\textwidth]{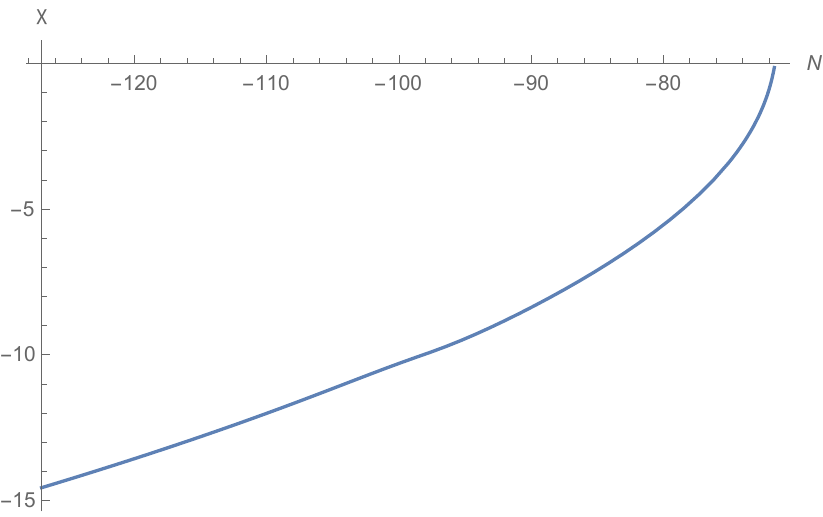}
\includegraphics[width=0.45\textwidth]{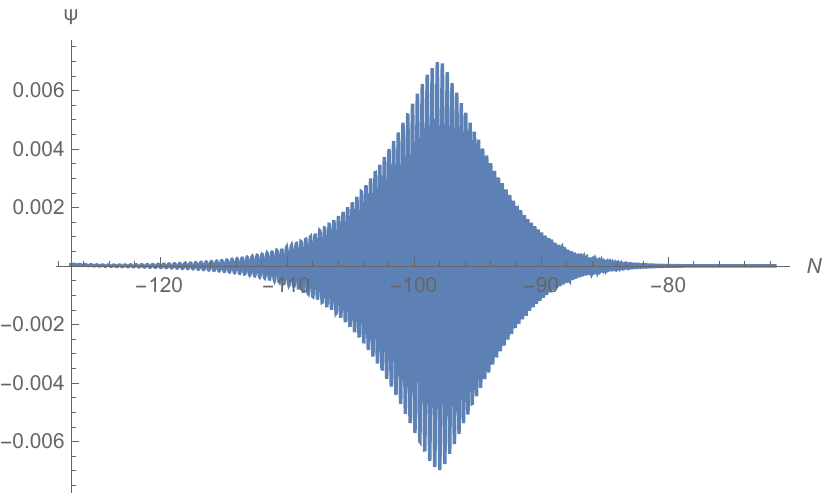}

\includegraphics[width=0.45\textwidth]{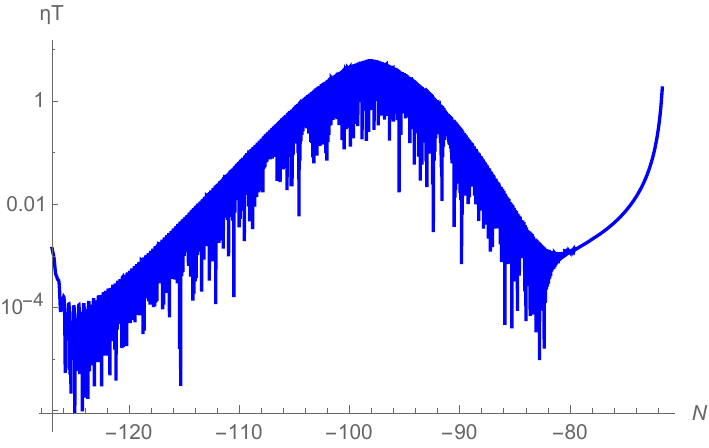}
\includegraphics[width=0.45\textwidth]{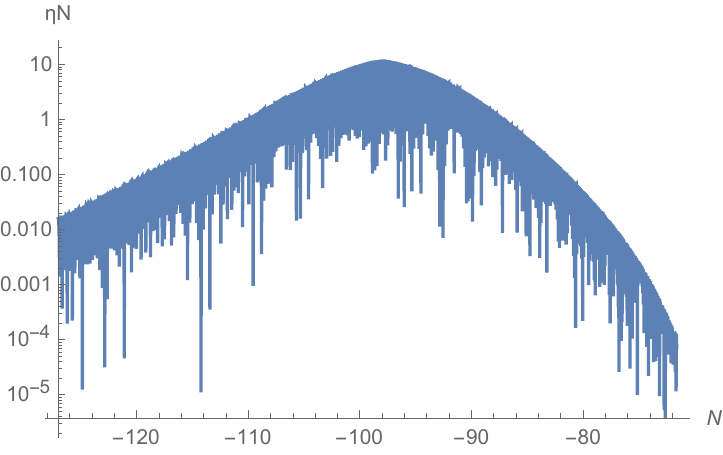}

\includegraphics[width=0.45\textwidth]{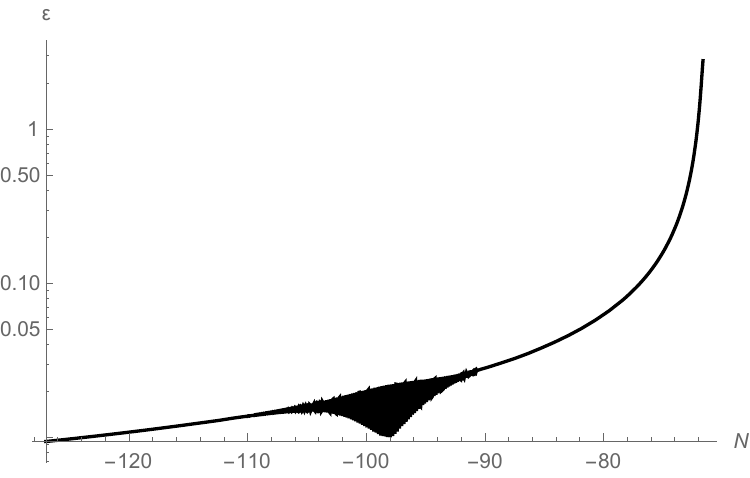}
\includegraphics[width=0.45\textwidth]{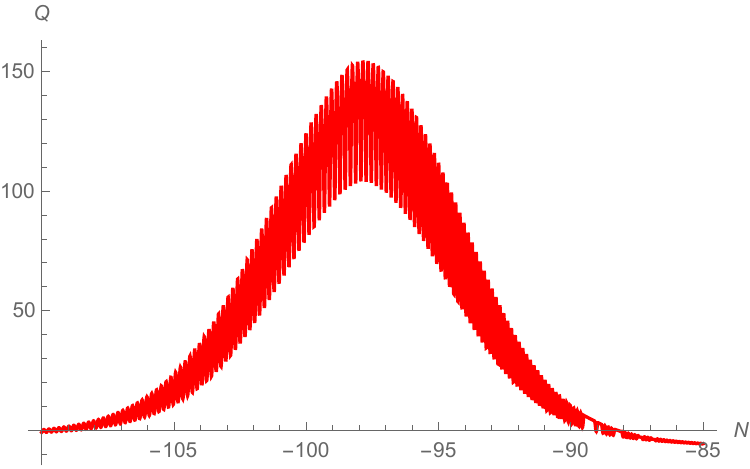}
\caption{
 For the model \eqref{equ:modelxsq} and \eqref{equ:modelxsq1}, we show the resulting numerical solution of the background evolution for: Up:  the scalar fields $\chi(N)$ (left) and $\psi(N)$ (right). Middle: the slow-roll parameters $|\eta_T(N)|$ (left) and $|\eta_N(N)|$ (right).  Down: the slow-roll parameter $\epsilon(N)$ (left) and $Q(N)$ (right).
}
\label{fig:ex4p1}
\end{figure}

\begin{figure}[t]
\centering
\includegraphics[width=0.5\textwidth]{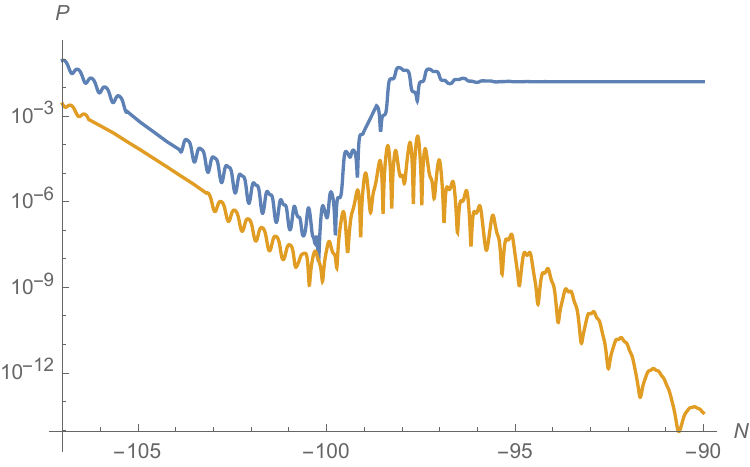}
\caption{For the model \eqref{equ:modelxsq} and \eqref{equ:modelxsq1}, for the Fourier mode $k=10^{-45} M_{PL}$, we plot $P_R(N)$ (blue) and $P_F(N)$ (orange).
}
\label{fig:ex4Pk}
\end{figure}

 We choose the following values for the parameters
\begin{equation}\label{equ:modelxsq1}
 m_{\chi}^2=2.3 \cdot 10^{-7} \ ,   \   m_{\psi}^2=1.2 \cdot 10^{-5} \ , \  \omega =100 \ ,   \  b=1.33 \cdot 10^{-5} \ , \  \rho=0.8 \ ,  \  \chi_*=-10 \  ,
\end{equation}
and we show the potential \eqref{equ:modelb1} to \eqref{equ:modelb3} in figure \ref{fig:potxsqtan}.
We integrate the equations of motion numerically from $N_i=-127$ to $N_f=-70$, with initial conditions $(\chi_{ini},\psi_{ini})=(-14.56317,0.00004)$ and $(\dot{\chi}_{ini},\dot{\psi}_{ini})=(0.13697, -0.00002)$ and present the results for the background evolution in figure \ref{fig:ex4p1}. We find a single, wiggly pulse of positive $Q$. Its peak height is $Q_{max} \approx 155$ but it remains positive for about $20$ efolds, in contrast to the alternating sign we found in our previous models. In figure \ref{fig:ex4Pk} we show  the evolution  of the $k=10^{-45}$ mode for $-107 \leq N \leq -90$.

\begin{figure}[t]
\centering
\includegraphics[width=0.45\textwidth]{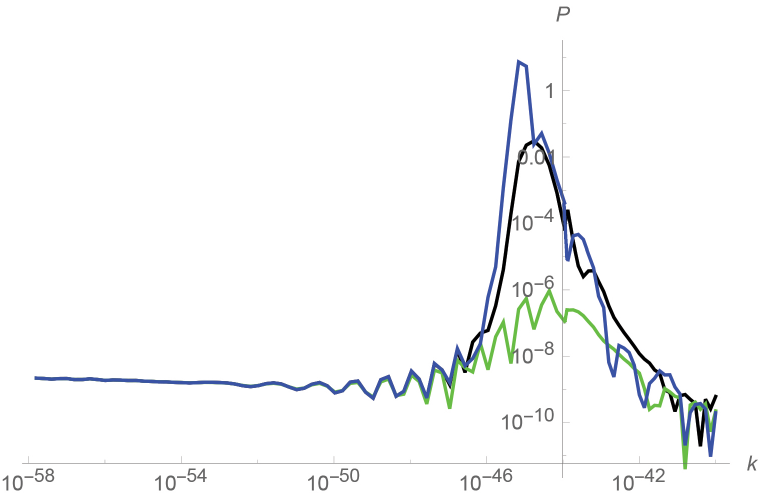}
\includegraphics[width=0.45\textwidth]{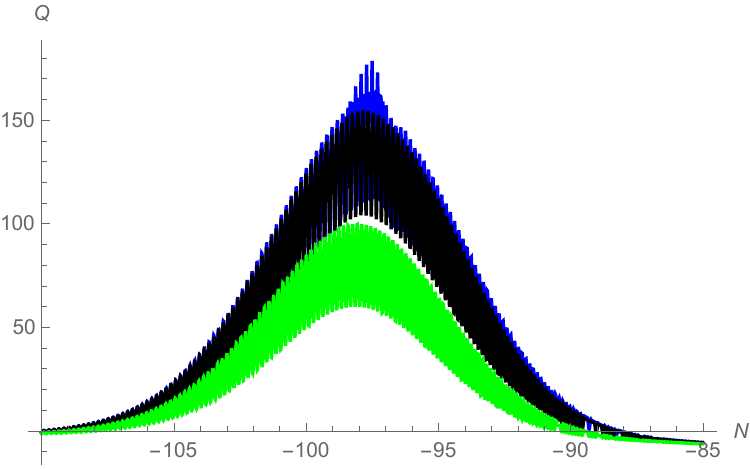}
\caption{For the model \eqref{equ:modelxsq} and \eqref{equ:modelxsq1} we plot, left: in black ($b=1.33 \cdot 10^{-5}$), the scalar power spectrum, $P_R$ for the band of Fourier modes $10^{-58}  \leq k \leq  10^{-40} $ at $N=-75$, when the modes have frozen; in green and in blue, the same quantity for $b= 1.1 \cdot 10^{-5}$  (green) and $b=1.342 \cdot 10^{-5}$ (blue). Right: the corresponding pulses of positive $Q$ for the three choices of $b$ with the corresponding colors.  
}
\label{fig:ex4Pk2}
\end{figure}

 At the left panel of figure \ref{fig:ex4Pk2} we show in black the power spectrum of the adiabatic mode for $10^{-58} \leq k \leq  10^{-40}$.  Like in the previous case, there are oscillations around the CMB amplitude for small wavenumbers while the very small wavenumbers remain unaffected, as expected. The enhancement peaks about $k= 10^{-45}$ ($7$ orders of magnitude) and is milder for smaller and larger scales.  We also show the corresponding power spectra for $b=1.342 \cdot 10^{-5}$ (blue) and $b= 1.1 \cdot 10^{-5}$ (green). At the right panel we show the pulses of positive $Q$ for each case. Like in section \ref{sec:enh}, we find that the precise choice of $b$ and the thickness of the envelope control the enhancement. For example, $b=1.1 \cdot 10^{-5}$ results into maximum enhancement by $3$ orders of magnitude, while $b=1.342 \cdot  10^{-5}$ results into maximum enhancement by $10$ orders of magnitude. In this model, the turns in field space are focused in a shorter timespan around $N_*$. As a result the bump in the power spectrum is thinner and more pronounced.

\begin{figure}[t]
\center
\includegraphics[width=0.4\textwidth]{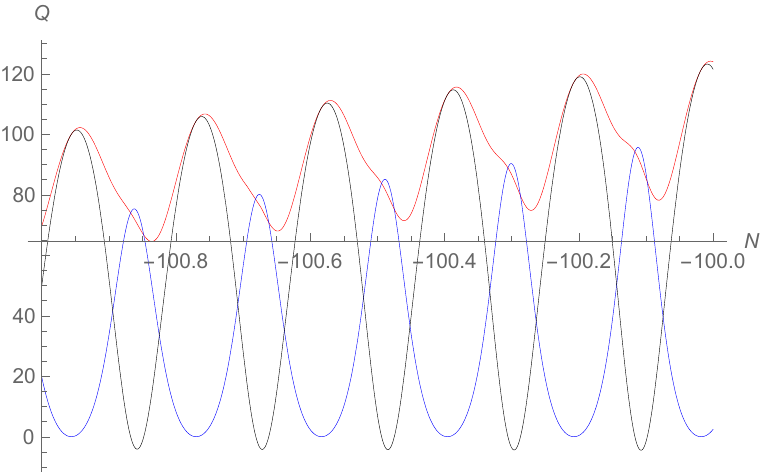}
\includegraphics[width=0.4\textwidth]{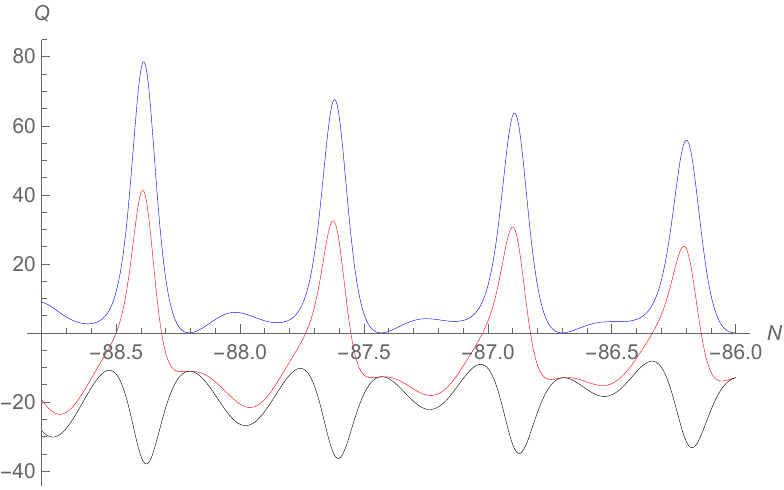}
\caption{For the model of section \ref{sec:part3}, we display in the left plot the pulses of positive $Q$ (red), as well as the contributions of the bending (blue) and $\mu^2$ (black) it comprises of, as described by equation  \eqref{equ:M}. The right plot is the corresponding figure for the model of section \ref{sec:sinu}.
}
\label{fig:dif2}
\end{figure}

\newpage

\section{Discussion}\label{sec:disc0}

\subsection{Features and limitations of valley types}\label{sec:new}

The effective mass for the entropic fluctuation receives three contributions. The first contribution comes from the centrifugal barrier of the turns in field space or equivalently the bending of the trajectory. The second contribution, $\mu^2$, is  the covariant Hessian of the potential along the isocurvarture or entropic direction. Potentials with certain convexity/concavity properties can thus contribute to pulses of positive $Q$ through $\mu$, as described by equation \eqref{equ:M}. Although not present in our model, the scalar curvature of the internal space of the fields also contributes when the fields have non-trivial internal geometry. In the left panel of figure \ref{fig:dif2} we show the balance of powers of these terms for the model of section \ref{sec:sinu}. The $\mu^2$ term is always a negative contribution that tends to stabilize the isocurvature mass, countering the effect of the bending,  decreasing or even erasing pulses of positive $Q$. The picture is different for the model of section \ref{sec:part3}, as we show in the right panel. Now the contribution of the $\mu^2$  term introduces additional, larger pulses of positive $Q$ between the peaks induced by the bending. As a result, several pulses of positive $Q$ merge into an extensive, wiggly pulse. We conclude that the $\mu^2$ term destabilizes the effective isocurvature mass in this case, thereby reinforcing the enhancement. 

The model with constant tilt can become sensitive to small changes of the parameters.
On the contrary, the profile of the oscillations in all plots of section \ref{sec:part3}, and consequently also the profile of the pulses in $Q$ and the evolution of the adiabatic amplitude, are much more canonical. The oscillations increase monotonically in height towards the peak of the feature and then decrease monotonically after it.  The frequency of the oscillations also displays a similar pattern. 

Even though we expect that our computation captures the main peak of the spectrum,  more sophisticated computational techniques are required to continue the evaluation for very large $k$.  The problem is that when we solve the coupled system \eqref{equ:flu} for the fluctuations, the computation cannot proceed if we insist on starting from $N_{ini}$ for the larger wavenumbers. We have to shift the initial time accordingly, which in turn makes the choice of initial conditions ambiguous.

\subsection{Reverse engineering primordial black holes and gravitational waves}\label{sec:adv}

Primordial gravitational waves can be sourced by scalar fluctuations since they couple at the second order in the Einstein's equations. The amplitude of the produced signal can in principle become detectable in the (near) future,  if the scalar fluctuations are somehow amplified by several orders of magnitude \cite{Caprini:2018mtu,Baumann:2007zm}. The authors of \cite{Ananda:2006af} considered the evolution equations for the second order tensor perturbations sourced by the scalar density perturbations and investigated how the system reacts to power being put in at one particular scale. They found that a delta function in the scalar spectrum indeed leads to a distinct peak in the tensor spectrum. Their findings also suggest that the scale where the tensor power spectrum is enhanced is roughly the scale where the feature was injected in the scalar spectrum.  The amplification in the tensor spectrum around the scale of the feature is comparable\footnote{This is reflected by their modulating function $\mathcal{F}$ being of order $0.1$ to $10$ around the peak.} in magnitude with the one  assumed for the scalar spectrum. The authors of \cite{Braglia:2020taf} studied how two-field models with  power spectra similar to those resulting from the sinusoidal valley can produce oscillations in the gravitational wave background. A proper calculation is necessary to make definitive statements for each case, but their analysis on $\Omega_{GW}$ is indicative of the results we expect. The peak in the scalar power spectrum formed by the assisted enhancement should analogously be reflected into a peak in the tensor spectrum. From that perspective, a peak injected by a sinusoidal valley at mHz frequencies could be relevant to the future observations of LISA, while a peak at nHz could be relevant to pulsar timing array observations. As far as the latter is concerned, recent reports \cite{NANOGrav:2023gor,EPTA:2023xxk,Ellis:2023oxs,EPTA:2023fyk,Reardon:2023gzh,Xu:2023wog} hint towards a stochastic background of cosmological origin, the deciphering and modeling of which has attracted great interest from the scientific community.

Black holes generated by primordial fluctuations enhanced by about 7 orders of magnitude are often discussed as a  possible solution to the dark matter problem. A rough dictionary between the assumed peak frequency of the bump in the power spectrum, the wavelength of the mode and the preferred mass of the resulting primordial black hole spectrum reads 
\begin{equation}
k\approx   6 \times 10^5 \frac{f}{10^{-9} \ \text{Hz}} \ \text{Mpc}^{-1}, \quad  M_H \approx 33 \left(\frac{10^{-9} \  \text{Hz}}{f}\right)^2  M_0 \ ,
\end{equation}
where $M_0$ is the solar mass. The preferred frequencies picked up by pulsar timing arrays are about the nHz territory and could be induced by amplified scalar fluctuations that will collapse into solar mass primordial black holes. LISA is planned to pick up mHz frequencies that could be induced by amplified scalar fluctuations that will collapse into $10^{-15} \leq  M_0 \leq  10^{-11}$ primordial black holes. The wide range of LISA actually covers the sublunar and asteroid range. Remarkably, these are three famous windows where primordial black hole abundance is much less constrained by observations and thus could be high enough to account for a large portion or even all of dark matter \cite{DeLuca:2020agl,Carr:2021bzv}. It would be interesting to explore whether our recipe can be utilized to produce viable models with signals observable by the gravitational wave detectors or the primordial black hole detection experiments.
 
Let us first discuss how to model which scales get enhanced. In section \ref{sec:enh2} we argued that increasing (decreasing) $N_*$ simply\footnote{To the precision level where  $H$ remains constant during the enhancement.} shifts the power spectrum to bigger (smaller) wavenumbers around the corresponding $k_{peak}$, while approximately preserving its shape. For the CMB scale we have 
$k_{CMB}= 0.05$ Mpc $^{-1}  \approx 5 \cdot 10^{-16}$ Hz $ \approx  10^{-58}  M_{PL}$. 
The scale where the enhancement peaks is 
$k_{peak}= 0.38 \cdot 10^P$ Mpc $^{-1}  \approx 4 \cdot 10^{P-15}$ Hz $ \approx  10^{P-57}  M_{PL}$. 
Using eq. \eqref{equ:peakenh}, we can write
\begin{equation}
 N_{*}-N_{CMB} \approx \ln(k_{peak}/k_{CMB}) \approx 2+2.3 P \ .
\end{equation}
According to this formula, a feature occurring about $16$ efolds later than the event of CMB crossing the horizon (such as the model of section \ref{sec:sinu}), would lead to a peak in the scalar power spectrum in nHz frequencies. Likewise, a feature at about $30$ efolds after CMB (such as the model of section \ref{sec:square}), would lead to a peak in the scalar power spectrum in mHz frequencies. We can arrange for that by shifting the parameter $\chi_0$ of the envelope of the sine-type interaction. The exact way in which $\chi_0$ controls $N_*$ is different for each model, but very easy to to figure out numerically in practice.

There are further observational constraints the model has to satisfy, such as the spectral index $n_s$ and the tensor-to-scalar ratio $n_T$. These are mostly\footnote{Some final tweaking might still be necessary.} taken care of by the the part of the Lagrangian that only features $\chi$, $L_{sf}(\chi)$, and thus can be borrowed to some extend from the single-field inflationary model. The coefficients of $L_{sf}(\chi)$ and the initial point of the trajectory $\chi_{ini}$ are predominantly responsible for $n_s$. Then the type of the potential, $\chi^2$, $\chi^3$, etc is responsible for $n_T$. 

Finally, for a given envelope $W$, the sinusoidal valley provides us with a knob that controls the amount of enhancement. This is the precision to which the value of $b$ satisfies the critical slow-down. $b_{crit}$ can be easily found numerically as the maximum value of $b$ for which the classical trajectory ends on the true minimum rather than on an intermediate, local minimum. Increasing the precision yields enhancement by additional orders of magnitude.

\section{Conclusions and Outlook}\label{sec:disc}

The sinusoidal valley was constructed to produce turns in field space by constraining the motion of background fields to a snaking path. Turns can become very sharp when the descend of the inflaton slows down almost completely during the turn. This is achieved by tuning the coupling $b$ of the sine-type interaction between the two scalars, and results into spikes in $\eta_N$ and thus $Q$, the precursor of the enhancement. A quick enough succession of pulses of positive $Q$ can greatly enhance the scalar amplitude in a step-by-step procedure. If an envelope is implemented on the sine-type interaction, the enhancement peaks at the scale that crosses the horizon when the central feature is crossed by the inflationary evolution and thus a bump is generated in the scalar power spectrum. Both the peak scale and the peak height can be controlled through the parameters of the inflaton potential.

 The phenomenological sine-type interaction can be incorporated in trivial generalizations of the standard inflationary models as an addition that simply injects the bump in the scalar power spectrum on top of their predictions.  We expect that such inflationary models can be tuned to satisfy the current observational constraints and also make interesting predictions for primordial black hole abundances and primordial gravitational waves. Detailed calculations for a concrete model will be addressed in future work.
 
 Despite the exotic look of this  interaction, sine-type potentials are not hard to find in particle physics. In fact they are commonly met as interactions with Goldstone bosons in  Lagrangians that describe physics below the energy scale where a global symmetry is spontaneously broken. Such a mechanism has been discussed in the literature in the context of natural inflation \cite{Freese:1990rb,Adams:1992bn,Stein:2021uge}, where the flatness of the inflaton potential is guaranteed by the shift symmetry of the resulting Goldstone boson. Axions are particles that can naturally play this role and their existence is on its own strongly motivated in theoretical particle physics for a number of reasons. For instance, inflationary models with axions have been thoroughly discussed in string theory in terms of axion monodromy inflation \cite{McAllister:2008hb,Flauger:2009ab}, with  potentials that include sines of the fields.  

It is more difficult to generate multifield interaction terms of the type $\psi\sin(\chi)$ through a top-down approach from string compactifications, successful moduli stabilization, etc. It would be very interesting to see whether such terms, and thus sinusoidal valleys, can appear in string theory. There is  an interesting recent study \cite{Bhattacharya:2022fze} of a supergravity model that exhibits a similar evolution. This model is a nice example of how the mechanism can occur in more fundamental scenarios of theoretical physics. In the light of our analysis, we believe that the source of the enhancement resides in the second term of their superpotentials that results into the sine-type axion-saxion term $\rho \sin{(b \theta)}$. The natural occurrence of a curvature term in supergravity scenarios can also contribute to the mechanism of assisted enhancement, as we mention in section \ref{sec:new}.

\appendix


\section{Background Evolution}\label{App:bg}

 The dynamics for the action \eqref{equ:action} of  the two scalar fields $\vec{\phi} =(\chi,\psi)$ 
are given by the following set of  equations of motion:

\begin{subequations}\label{equ:eomt}
\begin{align}
\label{equ:eomt:1}
\ddot\chi_0+3H \dot\chi_0+V_{,\chi}&=0
\\
\label{equ:eomt:2}
\ddot\psi_0+3H \dot\psi_0+V_{,\psi}&=0
\\
\label{equ:eomt:3}
 \sqrt{\frac{1}{6}(\dot\chi_0^2+\dot\psi_0^2)+\frac{V}{3}}&=H  \ ,
\end{align}
\end{subequations}
where $H= \dot{a}/a \ $ is the Hubble rate of expansion and we use the shorthand notation $V_{,\chi}$ and $V_{,\psi} \ $ for the partial derivatives $\partial V(\chi_0,\psi_0)/ \partial \chi_0 \ $ and $\partial V(\chi_0,\psi_0)/\partial \psi_0 \ $ respectively, evaluated for the background fields.  Combining the above equations of motions we also derive

\begin{equation}
\dot{H} = -\frac{1}{2}(\dot\chi_0^2+\dot\psi_0^2) \ .
\end{equation}
Once we supplement the equations of motion with an expression for $V$ and a set of initial conditions for the two scalar fields,
the equations of motion can be (numerically) integrated to yield a trajectory for the background evolution in field space $\chi-\psi$ as a function of cosmic time $t$. We use the equation of motion to substitute all second derivatives in favour of partial derivatives of the potential $V_{,\chi}$ and $V_{,\psi} \ $.
The slow-roll parameters that account for the (slow) variation of the background quantities during the inflationary evolution are defined as:

\begin{equation}
\epsilon = -\frac{\dot{H}}{H^2}=\frac{3(\dot\chi_0^2+\dot\psi_0^2)}{(\dot\chi_0^2+\dot\psi_0^2)+2V} \ ,
\end{equation}
and $\vec{\eta}=(\eta^{\chi},\eta^{\psi}) $ ,
where the components of the vector are given by

\begin{equation}
\eta^{\chi}=-\frac{\ddot\chi_0}{H \sqrt{\dot\chi_0^2+\dot\psi_0^2}} =\frac{ (3H \dot\chi_0+V_{,\chi})}{H \sqrt{\dot\chi_0^2+\dot\psi_0^2}}\ ,
\end{equation}

\begin{equation}
\eta^{\psi}=-\frac{\ddot\psi_0}{H \sqrt{\dot\chi_0^2+\dot\psi_0^2}} =\frac{ (3H \dot\psi_0+V_{,\psi})}{H \sqrt{\dot\chi_0^2+\dot\psi_0^2}}\ .
\end{equation}

So far we have been using the $\chi$ and $\psi$ axes to parametrize motion in the two-field space. We now rotate the axes and parametrize with respect to the directions tangent and normal to the trajectory in field space instead \cite{Gordon:2000hv}.  This choice could be beneficial, especially in cases where the motion is constrained by walls, and thus fluctuations perpendicular to the trajectory cost more energy.  The tangential orthogonal unit vector is $\vec{T}=(T^{\chi},T^{\psi})$, such that:
\begin{equation}
T^{\chi}=\frac{\dot{\chi_0}}{\sqrt{\dot\chi_0^2+\dot\psi_0^2}} \ , \quad
T^{\psi}=\frac{\dot{\psi_0}}{\sqrt{\dot\chi_0^2+\dot\psi_0^2}} \ .
\end{equation}
The normal orthogonal unit vector is $\vec{N}=(N^{\chi},N^{\psi})$, such that:
\begin{equation}
N^{\chi}=T^{\psi}\ , \quad N^{\psi}=-T^{\chi}\ .
\end{equation}
We can now decompose $\vec{\eta} \ $ into its projections along these newly defined orientations, $\vec{\eta}=\eta_{T}\vec{T}+\eta_{N}\vec{N} $,
to find

\begin{equation}
\eta_T= \vec{\eta} \cdot \vec{T}=-\frac{\ddot\chi_0\dot\chi_0+\ddot\psi_0\dot\psi_0}{H (\dot\chi_0^2+\dot\psi_0^2)}=3+\frac{ V_{,\chi}\dot\chi_0+ V_{,\psi}\dot\psi_0}{H (\dot\chi_0^2+\dot\psi_0^2)} \ , 
\end{equation}
and

\begin{equation}\label{equ:ap2}
\eta_N= \vec{\eta} \cdot \vec{N}=- \frac{\ddot\chi_0\dot\psi_0-\ddot\psi_0\dot\chi_0}{H(\dot\chi_0^2+\dot\psi_0^2)}= \frac{V_{,\chi}\dot\psi_0-V_{,\psi}\dot\chi_0}{H(\dot\chi_0^2+\dot\psi_0^2)}  \ .
\end{equation}
It is easy to see that 

\begin{equation}
\eta_T=\epsilon-\frac{\dot{\epsilon}}{2H\epsilon} \ ,
\end{equation}
is equivalent to the standard second slow-roll parameter of single-field inflation. On the other hand, $\eta_N$ parametrizes the rate of turn of the trajectory in field space. This can be deduced from the equations governing the evolution of the unit vectors $\vec{N}$ and $\vec{T}$:

\begin{equation}
\dot{\vec{T}}=-H\eta_N \vec{N} \ , \quad
\dot{\vec{N}}= H\eta_N \vec{T} \ .
\end{equation}

For most of our analysis it will be convenient to measure time in terms of efolds of expansion, $dN=H(t)dt$.
This can be easily done by replacing all derivatives with respect to time with  derivatives with respect to the efolds times the Hubble expansion, $
(\ )_{,t} \to H(N) \cdot ( \ )_{,N}$  .  For notational simplicity, we will also drop the zero subscript from $\psi$ and $\chi$, although we will be still referring to the background solutions. We then arrive at the equations of motion \eqref{equ:eom}
and the expansion parameters \eqref{equ:eps} to \eqref{equ:etan}.

\section{Qualitative features}\label{App:anal}

\begin{figure}[t]
\centering
\includegraphics[width=0.6\textwidth]{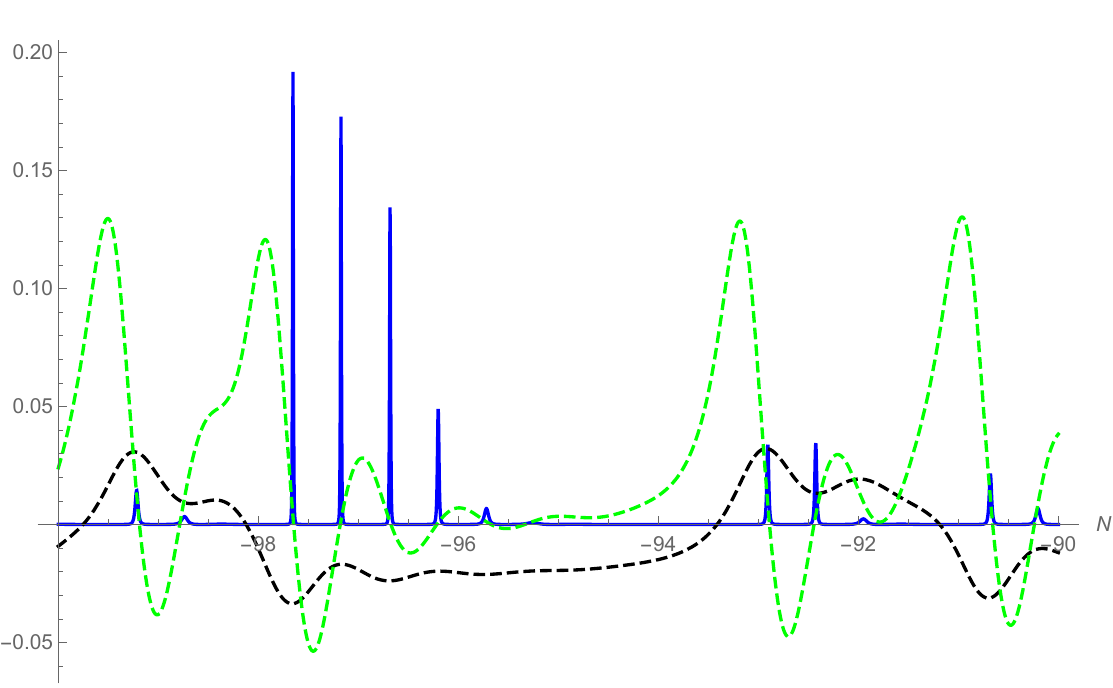}
\caption{
For the model of section \ref{sec:sinu} we show the peaks of the numerical solution for $\eta_N^2/10^5$ (in blue) plotted over the  numerical solution for $\psi$ (black-dashed) and the numerical solution of $\chi_{,N}$ (green-dashed). We notice large peaks at $t_{\alpha}$ for $\eta_N^2$ while also $\chi_{,N}(t_{\alpha})=0$.
}
\label{fig:exp1}
\end{figure}

The analytical approximation of this section is based on our expectation that the classical motion of the inflaton will be constrained to the flat direction of the potential, the sinusoidal valley. We obtain the valley trajectory by demanding that the $\psi$-derivative of  eq. \eqref{equ:pot1} vanishes.\footnote{If we also minimize the potential in the $\chi$-direction we would obtain the minimum, where the inflaton sits at the end of inflation. The potential \eqref{equ:pot1} can be modified to accommodate a such minimum.}

\begin{equation}\label{equ:val}
\psi = \frac{b}{2 m^2} \sin(\omega \chi)  \ .
\end{equation}
The motion in $\psi$-axis resembles that of a pendulum.
For simplicity, the analysis of this section will be carried out in cosmic time $t$. Then, for the rest of the paper we will switch back to efolds. Being a mere rescaling of the time axis, this change of time variable  cannot invalidate any of our qualitative conclusions.
We use the valley trajectory to eliminate any $\psi$-dependence in the expressions \eqref{equ:etan} and \eqref{equ:M} for $\eta_N$ and $(\mu/H)^2$ respectively:

\begin{equation}\label{equ:etaN2}
\eta_N =  -\frac{b \omega}{8 m^2} \ \frac{4 c m^2 + b^2 \omega \sin(2 \omega \chi)}{H m^2  \left(1+\left(\frac{b \omega}{2 m^2}\cos( \omega \chi)\right)^2\right)} \  \frac{\cos(\omega \chi)}{\dot{\chi}} \ ,
\end{equation}

\begin{equation}\label{equ:MoH2}
(\mu/H)^2 = \frac{1}{H^2 m^2 \left(1+\left(\frac{b \omega}{2 m^2}\cos( \omega \chi)\right)^2\right)}{\left( 2m^4 +(b \omega \cos(\omega \chi))^2+2\left(\frac{b^2 \omega^2 \sin (2 \omega \chi)}{8 m^2}\right)^2   \right)} \ .
\end{equation}

From eq. \eqref{equ:etaN2} we expect large values for $\eta_N$ when the $\chi$-axis motion slows down to a near stop, that is, at $t_{\alpha}$, such that $\dot{\chi}(t_{\alpha}) \approx 0$. This is confirmed by the numerical calculation as we show in figure \ref{fig:exp1}. In fact, since $\dot{\chi}$ becomes negative, it has additional roots at $\tilde{t}_{\alpha}$, which spawn further peaks in $\eta_N$. 

We  now use the valley constraint to eliminate any $\psi$-dependence from the equations of motion in favor of $\chi$:

\begin{subequations}\label{equ:EOM1}
\begin{align}
\label{equ:EOM1:b}
\ddot\chi+3H \dot\chi - c -\frac{b^2 \omega}{4 m^2}\sin(2 \omega \chi)=0 
\\
\label{equ:EOM1:c}
H= \frac{1}{2\sqrt{6}}\sqrt{8 V_0 -\frac{2b^2}{m^2}\sin(\omega \chi)^2 - 8 c \chi +\left(4+\frac{b^2 \omega^2}{m^4}\cos(\omega \chi)^2 \right)\dot{\chi}^2} \ .
\end{align}
\end{subequations}
For the interesting part of the parameter space for pulses of positive $Q$, where we require points with $\dot\chi\approx 0$. The oscillations together with the shallow wells slow down the descend of the inflaton. The right choice of parameters should bring the local minima of $\dot{\chi}$ close to zero or even at negative values like the case we show in figure \ref{fig:exp1}. Indeed, one can confirm numerically from eq. \eqref{equ:EOM1} that the ball stops traveling in the $\chi$-axis when $\frac{b^2 \omega}{4m^2}>c$. On the other hand, for $\frac{b^2 \omega}{4m^2}<c$ the minimum velocity obtained when traversing the max-$\psi$ displacement points increases and thus the ball does not slow down enough for pulses of positive $Q$ to take place. We arrive at the approximate formula  
 \begin{equation}\label{equ:chicond1}
 \frac{b^2 \omega}{4m^2} =  c  - \delta_1\ ,
 \end{equation}
which we call `critical slow-down' approximation. This $\delta_1$ is a very small number. 
We will use this as an indicator for $b$:

\begin{equation}
b = 2m \sqrt{\frac{( c -\delta_1)  }{ \omega}}  \ ,
\end{equation}
for the numerical calculations, or substitute

\begin{equation}\label{equ:csd2}
b  \approx 2 m\sqrt{\frac{c  }{ \omega}} \ ,
\end{equation}
 in the analytic approximations.

Upon applying the valley approximation \eqref{equ:val} and the critical slow-down \eqref{equ:csd2}, $Q$ becomes:

$$ Q = \frac{1}{H^2\left(1+J\cos( \omega \chi) \right)^2} \times $$

\begin{equation}\label{equ:analQ}
\times  \left[\frac{1}{\dot\chi^2} \frac{c^2 J \cos(\omega \chi)^2\left(1+\sin(2\omega \chi)\right)^2}{ 1+J\cos(\omega \chi)^2}-2\Big(1+2J\cos(\omega \chi)^2+\frac{J^2}{4} \sin(2\omega\chi)^2\Big) \right]
\end{equation}
where we have defined $J \equiv ( c \omega) /m^2$ .  Let us inspect eq. \eqref{equ:analQ}. If we ignore the role of  $ \dot\chi^2$ for now,  the negative term in the square brackets wins in the limit of small and large $J$. This means that we cannot increase $Q$ by going to extreme values of $J$, $\omega$ or $m^2$. The way to obtain pulses of large $Q$ is to make $1/\dot\chi(t)$ very large. This is exactly what the critical slow-down constraint does: it brings $\dot\chi(t)$ close to the $\dot\chi=0$ axis.

Very roughly, we can approximate \eqref{equ:EOM1:b} around the maximum of the sine as
\begin{equation}
\ddot\chi+3H \dot\chi -\left( c -\frac{b^2 \omega}{4 m^2}\right)=0  \ .
\end{equation}
At $b_{crit}$ the solution is $x \sim \exp{(-t)}$, while away from it it develops a dominant part $\sim t$. This indicates that $1/\chi'$ could be  exponentially suppressed when we move away from $b_{crit}$. In the light of our previous discussion of \eqref{equ:analQ}, it also implies  that $Q$ is exponentially suppressed away from $b_{crit}$, 
\begin{equation}\label{equ:expb}
Q \sim Q_0\exp\left(-{c_1|b_{crit}-b|/b_{crit}}\right) \ ,
\end{equation}
where $Q_0$ is the maximum of the positive-Q pulse and $c_1$ a constant.

We have confirmed numerically that pulses of positive $Q$  usually require $0.2<J<20$. This constrains $\omega$ once $c$ and $m$ have been chosen. We have also confirmed that values for $b$ that satisfy the critical slow-down more precisely enhance $Q$ significantly. 

\acknowledgments

I would like to thank N. Tetradis for our useful discussions.



\end{document}